\documentclass[twocolumn]{aastex701}
\usepackage{hyperref}
\usepackage{amsmath}
\usepackage{amssymb}
\usepackage{tabularx}
\usepackage{color, colortbl}
\usepackage{hhline}
\usepackage{multirow}
\usepackage{makecell}
\usepackage{booktabs}
\usepackage{threeparttable}

\newcommand{\UChicago}{\affiliation{Department of Physics, The University of Chicago, 5640 South Ellis Avenue, Chicago, Illinois 60637, USA}}

\newcommand{\KICP}{\affiliation{Kavli Institute for Cosmological Physics, The University of Chicago, 5640 South Ellis Avenue, Chicago, Illinois 60637, USA}}
\newcommand{\EFI}{\affiliation{Enrico Fermi Institute, The University of Chicago, 933 East 56th Street, Chicago, Illinois 60637, USA}}
\newcommand{\UChicagoAA}{\affiliation{Department of Astronomy and Astrophysics, The University of Chicago, 5640 South Ellis Avenue, Chicago, Illinois 60637, USA}}

\usepackage{graphicx}
\usepackage{amsmath}
\usepackage{amssymb}
\usepackage[T1]{fontenc}
\defcitealias{EHT1}{EHTC I}
\defcitealias{EHT2}{EHTC II}
\defcitealias{EHT3}{EHTC III}
\defcitealias{EHT4}{EHTC IV}
\defcitealias{EHT5}{EHTC V}
\defcitealias{EHT6}{EHTC VI}
\defcitealias{1997ASPC..125...77S}{Shepherd 1997} 
\defcitealias{2011ascl.soft03001S}{Shepherd 2011}
\defcitealias{PRIMO23}{Medeiros et al. 2023a}

\received{February 24, 2026}
\revised{June 22, 2026}
\accepted{July 15, 2026}

\begin{document}

\title{A simple model for extracting astrophysics from black hole images}

\author[0000-0002-8304-0109]{Alexandra G. Guerrero}
\email[show]{aghanselman@uchicago.edu}
\UChicago

\author[0000-0002-0175-5064]{Daniel E. Holz}
\email{holz@uchicago.edu}
\UChicago
\EFI
\UChicagoAA
\KICP

\begin{abstract}
The Event Horizon Telescope (EHT) is providing unprecedented high-resolution images of supermassive black holes. These images are fundamentally related to properties of the luminous accretion disks, since black holes themselves produce no light. We develop a simple prescription to relate observational black hole image features to a toy model for the intensity profile of the associated accretion disk. We apply our model to both the original EHT image of M87* and the reanalyzed image from the $\texttt{PRIMO}$ algorithm, providing generic, simultaneous constraints on the black hole mass and disk emission properties. While current images lack the resolution to confidently detect the photon ring, we use multiple observed features from the original EHT image to constrain M87*'s mass to $6.6^{+1.2}_{-1.0}\times 10^9 M_\odot$, and find emission may extend to the black hole horizon. Conversely, using constraints from the $\texttt{PRIMO}$ image alongside brightness asymmetry constraints from the original EHT analysis yields a mass of $6.4^{+0.7}_{-0.7}\times 10^9 M_\odot$, with the disk’s inner edge between $3M$ and $5.3M$. Both analyses rule out the disk’s inner edge coinciding with the innermost stable circular orbit for a Schwarzschild black hole ($6M$), and our analysis with $\texttt{PRIMO}$ confidently rules out significant emission extending to the horizon ($2M$). While this work restricts disks to Keplerian orbits outside the ISCO, we demonstrate that brightness asymmetry is especially sensitive to the disk velocity profile. Further assumptions on the mass of M87* and connections between the accretion disk cutoff and physical radii allow for rudimentary black hole spin estimates.
\end{abstract}

\section{Introduction}
The Event Horizon Telescope (EHT) collaboration released the first image of a black hole in the center of the galaxy M87 (hereby denoted M87*)~\citep[][; hereafter EHTC I\mbox{--}VI]{EHT1, EHT2, EHT3, EHT4, EHT5, EHT6}, and has subsequently released an image of the black hole at the center of our own galaxy (denoted Sgr A*)~\citep{EHTSgr1,EHTSgr2,EHTSgr3,EHTSgr4,EHTSgr5,EHTSgr6}. Further observations of polarization of M87* have found evidence for the presence of magnetic fields~\citep{2021ApJ...910L..12E, 2021ApJ...910L..13E, EHT9}, as well as demonstrated the persistence of the features of the bright central ring~\citep{2024A&A...681A..79E,2025A&A...693A.265E}. 

These observations have opened a new avenue to investigate the strong-gravity regime and the astrophysical environments surrounding supermassive black holes at the centers of galaxies. One immediate example is EHT's estimation for the mass of M87*, finding a value of $\sim 6.5\times10^9\,M_\odot$~\citepalias{EHT6}; this is consistent with $\sim6.6\times10^9\,M_\odot$ attained from stellar dynamics mass estimates~\citep{StellarDyn}, but is inconsistent with the gas dynamics estimate of $\sim3.5\times10^9\,M_\odot$ from~\cite{GasDyn}. We note that more recent gas dynamics estimates agree with the reported mass from stellar dynamics~\citep[][see also \cite{Liepold2023} and \cite{2024MNRAS.527.2341S} for alternative dynamics mass measurements]{2023A&A...679A..37O}. Since the release of EHT's first images, there have been many studies demonstrating a wealth of scientific insight. This includes works studying how black hole images depend on astrophysics~\citep{Lia18,Chael21,LiaAsym22,Vincent21,Vincent22,2022ApJ...941...88O,2024ApJ...971..172T,2025arXiv250510333D}, determining accretion disk physics~\citep{Narayan19,2019MNRAS.486.2873C,White20,2025ApJ...980..262D,2025MNRAS.537.2496C}, and testing general relativity with black hole images ~\citep[e.g.][]{2019GReGr..51..137P,Gralla2020,Gralla2005,2020PhRvL.125n1104P,Lara21,2021CQGra..38uLT01V,2021PhRvD.103b4023G,2021PhRvD.103j4047K,2023ApJ...942...47Y}.

EHT uses very-long-baseline interferometry (VLBI) to probe event horizon-scale features in the frequency domain~\citepalias{EHT1,EHT2,EHT3}. EHT then synthesizes images using either geometric, ``generalized-crescent'' models or snapshots from general-relativistic mangetohydrodynamics (GRMHD) simulations and fits these images to VLBI data~\citepalias{EHT6}. Alternatively, images are produced directly through image reconstruction algorithms~\citepalias{EHT4}. Physical information about the black hole, such as its mass, is then extracted by calibrating the images to snapshots from a GRMHD image library~\citepalias{EHT5,EHT6}.

GRMHD simulations are extremely sophisticated, incorporating a wide dynamic range and diverse physical processes, making them generally computationally expensive\footnote{The Simulation Library used in EHT analyses takes up 23 TB of storage~\citepalias{EHT5}, and the state-of-the-art \texttt{H-AMR} code, for example, uses approximately $10^{13}$ cycles/GPU for a geometrically thick accretion disk with no adaptive mesh refinement~\citep{HAMR22}.}. Given the computational expense of these simulations, EHT investigated a restricted subset of initial conditions in the GRMHD simulations that compose their image library, focusing mostly on varying black hole spin and magnetic flux to investigate both `Standard and Normal Evolution'~\cite[SANE;][]{Narayan2012, Sadowski2013} and `Magnetically Arrested Disk'~\cite[MAD;][]{Igumenshchev2003, Narayan2003} accretion flows~\citepalias{EHT5}. However, certain simplifying assumptions are made, such as treating the plasma as an ideal fluid. Additional considerations, such as tilted disks or radiative cooling, were not included in their first analysis~\citepalias{EHT5}. It is to be noted that there has subsequently been a large body of work incorporating these additional physical prescriptions into GRMHD simulations~\citep[see e.g.][or \cite{patoka,Mizuno22} for a review]{2019MNRAS.486.2873C,White20,2022ApJ...935L...1L,HAMR22,2023MNRAS.518.3441C,2023ApJ...955...47S,2025MNRAS.537.2496C,2025ApJ...980..262D}. Black hole images are fundamentally related to properties of the accretion disk~\citep{1973ApJ...183..237C,Luminet1979,Broderick2009,Broderick2011,Broderick2016,Pu2016,Pu2018,GHW,EHT5,Chael21,Vincent22}, and therefore any information reported for M87* using the GRHMD image library can only be valid for the range of GRMHD simulations and assumptions considered in the analysis. 

Due to the complicated and theoretically uncertain nature of accretion disks ~\citep[see, e.g.][for a review of popular accretion disk models]{AbramowiczReview13}, it is prudent to include robust disk properties in image analyses. Recently, \cite{Palumbo22} developed a model using a simplified accretion disk model to constrain the black hole mass, spin, and inclination in black hole images. This model was tested on GRMHD data and found that spin is mostly unconstrained, with large degeneracies between accretion disk properties and astrophysical parameters. \cite{Chang24} extends this work, instead using a simplifed disk model assuming the emission comes from a bi-conical region to simulate that the majority of emission stems from the jet. Along with comparing to GRMHD images, \cite{Chang24} use this model to constrain astrophysical features using the 2017 EHT observations of M87*. While some astrophysical features, such as mass and inclination angle, are well constrained, others such as spin are still uncertain.

In this paper we present a computationally-inexpensive two-parameter toy model for the accretion disk emission profile which captures broad features of the intensity of the disk. Our simple toy model allows us to place more general constraints on emission profile features along with other astrophysical parameters, such as the mass of the black hole, and does not require the use of GRMHD simulations for calibration. 

Furthermore, since the release of the original EHT M87* images, there have been reanalyses of the data \citep[see e.g.][]{2107, 2208, PRIMO23} using more informed approaches, which provide updated estimates of observed image features. Along with the original EHT analysis presented in \citetalias{EHT4}, we use the updated image features reported in~\cite{PRIMO23} using the \texttt{PRIMO} algorithm~\citep{PRIMOintro} to constrain astrophysical quantities from the M87* images. \texttt{PRIMO} uses principal component analysis (PCA) on a large suite of GRMHD simulations as training data to reconstruct the image of M87$^*$ from the 2017 EHT data~\citep{PRIMO23}. Therefore, while our analysis on the original EHT M87$^*$ image is  agnostic to GRMHD simulations, our analysis comparing to the image generated with \texttt{PRIMO} will be informed implicitly by the GRMHD simulations used in \texttt{PRIMO}'s training set. Comparing to both the original EHT and the \texttt{PRIMO} images demonstrates how uncertainties in observed image features, as well as any GRMHD assumptions, influence resulting astrophysical constraints.

This paper is organized as follows: In \S \ref{sec:theory} we introduce our formalism  to generate theoretical black hole images. In \S\ref{sec:imageFeatures} we relate theoretical and observed image features, and introduce a method for extracting black hole masses and emission profile constraints. We then compare theoretical images to EHT's original image of M87* in \S\ref{sec:M87-Gauss}, and to the reanalyzed image from the \texttt{PRIMO} algorithm in \S\ref{sec:PRIMOresults}. Finally, we discuss the implications of our analysis in \S\ref{sec:implications}, and conclude in \S\ref{sec:discConc}.

\section{\label{sec:theory} Generating black hole images}
In this work we focus on the case of a central non-spinning supermassive Schwarzschild black hole. \cite{1910} demonstrated that for direct emission near the central black hole emitted at some radius $r$, the emission reaches a distant observer at some impact parameter $b$ such that $b/M \sim r/M + 1$, termed the ``just add one'' approximation. Extending this estimation to Kerr black holes results in $\lesssim 10\%$ variations in the ``just add one'' approximation for face-on disks for all spin values~\citep{1910}, and $\lesssim 20\%$ for any spin and inclination~\citep{2107}. Due to additional uncertainties in our analysis, such as blurring the generated images to match the instrumental resolution of the nominal 2017 EHT array \citepalias[see Appendix~G of][]{EHT4}, restricting our attention to Schwarzschild black holes is sufficient for this first analysis. To further confirm this choice, we demonstrate in Appendix~\ref{sec:SpinInvestigation} that, unlike the brightness asymmetry, angular-scale image features such as the diameter or width of the bright ring seen in EHT images are largely unaffected by adding spin to the central black hole, validating the use of these image features without incorporating a full Kerr analysis\footnote{Note, however, that Appendix~\ref{sec:SpinInvestigation} also demonstrates that the brightness asymmetry of the bright ring in black hole images does substantially change as a function of spin for accretion disk emission at radii sufficiently close to the event horizon. Therefore, care must be taken when using brightness asymmetry to determine any constraints on spin, or other astrophysical processes such as the accretion disk velocity profile.}$^,$\footnote{In Schwarzschild black holes, the accretion disk will always be equatorial. However, non-equatorial accretion disks are possible around a Kerr black hole if the disk is misaligned with the black hole's spin axis. In fact, tilted disks may be common around AGN~\citep[see e.g.][]{2024arXiv240410052F,2023Natur.621..711C}. Studies have shown that non-equatorial emission may change the size and structure of the resulting black hole image~\citep[see e.g.][]{2020MNRAS.499..362C,White20}. While we do not investigate misaligned disks in this work, analyses extending to Kerr black holes should include disk misalignment to prevent any systematics due to the effect of tilt on image parameters.}. In the following subsection, we first review geodesics in Schwarzschild spacetime before describing our toy accretion disk model and image generation technique in subsequent subsections.
\begin{figure}
    \centering
    \includegraphics[width = 0.8 \linewidth]{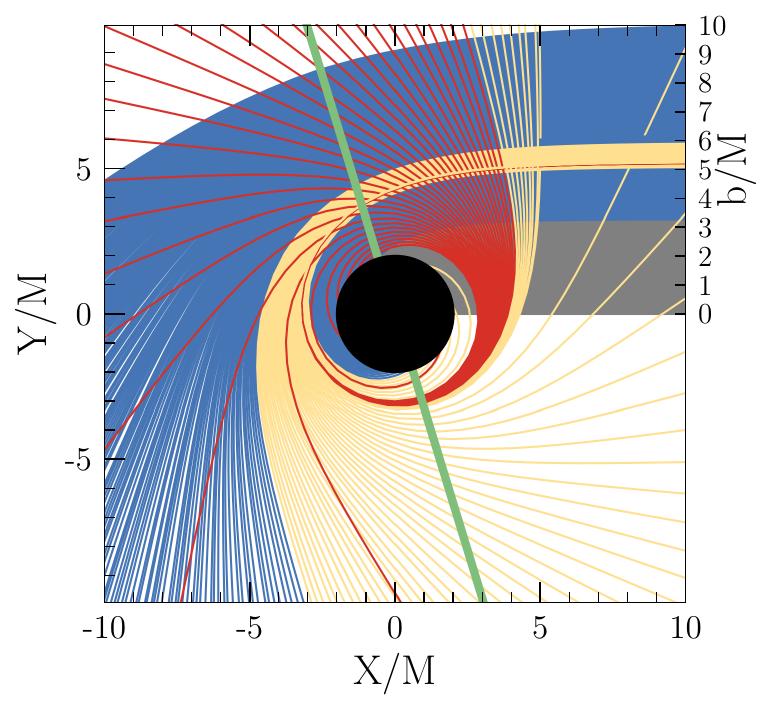}
    \caption{Two-dimensional cross-section of photon trajectories traced back from a distant observer located to the far right for the image-plane slice of $\alpha = \pi/2$. The left and bottom axes are in Schwarzschild radial coordinates, while the top right are in impact parameter coordinates. The black circle indicates the event horizon of the black hole and the green line indicates an accretion disk inclined at an angle $\theta_0 = 17^\circ$. Grey lines indicate the photons that never hit the accretion disk, while blue lines are the direct emission from the front of the disk, hitting the disk once. Yellow lines originate from the back of the disk and hit the disk twice. Red lines indicate the photon ring and hit the disk 3 or more times.}
    \label{fig:geodesics}
\end{figure}

\subsection{\label{sec:theoryBend}Review of Schwarzschild geodesics}
We follow the approach presented in \citet{GHW} to generate images. Recall that photons in a Schwarzschild spacetime travel on null geodesics with conserved energy $E$ and conserved angular momentum $L$. When considering photon trajectories as seen by a distant observer, only the ratio $b = L/E$ is relevant, where $b$ is the impact parameter~\citep{Hagihara1931,Darwin1959, Luminet1979}.  To generate an image as seen by this distant observer, we ray-trace null geodesics backwards from infinity and record the intersections these rays make with the accretion disk as they are lensed around the black hole. We parametrize the image plane of the distant observer in polar coordinates $(b,\alpha)$, where $b$ is the impact parameter and $\alpha=0$ is aligned to be perpendicular to the accretion disk tilt axis projected onto the image plane. We then generate `transfer functions' that relate the image plane coordinates $(b,\alpha)$ to the radial points $r_i$ where the rays hit the accretion disk, where the subscript $i$ is the $i$-th intersection of the disk. Figure~\ref{fig:geodesics} depicts a two-dimensional cross-section of rays that are traced back from a distant observer located to the far right of the plot for an impact parameter range of 0 to 10$M$ with $\alpha = \pi/2$. Following terminology from~\citet{GHW}, the $i = 1$ intersection (blue lines) corresponds to the direct image coming from the front of the disk, the $i = 2$ intersections (yellow lines) make up the ``lensing ring'' corresponding to emission from the backside of the disk, and the $i = 3$ intersection (red lines) corresponds to the ``photon ring.'' Emission for $i > 3$ corresponds to rays that asymptote to the critical curve, which for Schwarzschild black holes is located at $r = 3M$ corresponding to an impact parameter $b_c = 3 \sqrt{3} M$. The grey lines in Figure~\ref{fig:geodesics} correspond to rays that never hit the disk. As shown in \cite{Vincent22}, this ``central brightness depression'' only corresponds to the black hole ``shadow''---where intensity falls off inside the critical curve \citep[see e.g.][]{Bardeen1973, Luminet1979,Falcke2000, GHW}---when accretion is fully spherical and infalling~\citep[see also][]{Narayan19}. In our thin disk model, this central dark region need not correspond to the black hole shadow, and thus, we will use the common terminology that the central dark region in black hole images is the ``central brightness depression'' to distinguish our generic images from the special instance when this dark region aligns with the black hole shadow.

\subsection{\label{sec:theoryEm} Accretion disk properties}
As mentioned previously, light from black hole images originates from the accretion disk surrounding the black hole. For simplicity we take the accretion disk to be geometrically and optically thin and inclined at an angle $\theta_0$ with respect to a distant observer, where $\theta_0=0$ indicates a face-on disk. Although the accretion disk surrounding M87* is often considered to be an optically thin but geometrically thick disk~\citepalias{EHT5}, \cite{Gralla2020} and references therein show that using the ``equatorial approximation''---where the accretion disk is assumed to be geometrically thin---gives qualitatively good agreement to images with geometrically thick disks\footnote{\cite{Gralla2020} demonstrate that the photon ring is brighter in the case of geometrically thick disks. However, since we only consider blurred images that do not have the resolution to observe the photon ring (See \S\ref{sec:blurring}), using the equatorial approximation is sufficient for our purposes.}.

We assume the intensity profile, $I_{\rm em}$, of the accretion disk depends only on the radial position\footnote{Although we do expect some angular variation in the accretion disk~\citep{Conroy23}, our toy model should be sufficient to estimate the general properties of the subsequent images.} such that $I_{\rm em} = I(r)$. Since we wish to analyze the phenomenological characteristics of black hole images, we use a two-parameter toy model for the emission corresponding to an inverse power-law in intensity with a sharp inner cutoff:
\begin{eqnarray}
I_{\rm{em}} = I(r) \propto 
 \begin{cases}
      r^{\beta}, \ \rm{if} \ r \geq r_{\rm cut}\\
      0, \ \rm{if} \ r < r_{\rm cut} \; ,
    \end{cases}
    \label{eq:Ir}
\end{eqnarray}
with $\beta < 0$. A cutoff of $r_{\rm cut}=2M$ indicates the accretion disk extends all the way to the horizon, whereas if $r_{\rm cut} > 2M$ then the disk intensity cuts off before the disk reaches the horizon. This toy model allows us to probe a large range of possible images by varying emission profile parameters. Although a sharp cutoff is unlikely to be physical\footnote{In fact, \cite{2022ApJ...941...88O} argue that there cannot be a sharp cutoff due to conservation laws.}, it is expected to be a reasonable approximation to the overall emission structure for a disk-like accretion structure. For example, \citet{2022ApJ...941...88O} investigate image cross-sections for a toy model in disk emissivity that includes an arbitrary scale height, and investigates having a potential inner disk cutoff. As seen in Figure 4 of \citet{2022ApJ...941...88O}, introducing a sharp cutoff at radii outside the horizon introduces little change for a thin disk in terms of the generated image, except for a slight difference in how sharp the image intensity decays inside the photon ring (e.g. $X/M \lesssim 5M$ for a horizontal cross-section). Since the central brightness depression ratio is not well constrained~\citepalias{EHT4}, this difference in the amount of intensity inside the central brightness depression cannot be constrained with current observations. For disks that cut off outside the critical curve, there will be a gap between the photon ring and the direct emission spectrum. This gap would be clear evidence for a distant disk cutoff.
However, given the resolution of the instruments, this gap may not be resolved. Therefore, we stress that our model including a sharp cutoff still generates images that are consistent with current observations. The main benefit of introducing a sharp cutoff is to introduce the flexibility to infer any potentially unexpected features, like a disk that cuts off before reaching the horizon.

\subsection{Simulating observed images}\label{sec:blurring}
To generate the observed images, we first consider from Liouville's Theorem the fact that $I/\nu^3$ is conserved along the ray, where $\nu$ is the frequency of the light ray. Then, $I_{\nu^\prime} = (\nu^\prime/\nu)^3 I(r) = g^3 I(r)$, where $I_\nu$ is the specific intensity in the disk's stationary frame, $I_{\nu^\prime}$ is the specific intensity observed by a distant observer, and $g = \nu^\prime/\nu$ is the redshift factor. If the disk is stationary, the gravitational redshift around a Schwarzschild black hole is given by $g = \sqrt{1 - 2M/r}$. If the disk is instead rotating about the black hole, there is an extra redshift factor due to the Doppler effect. We follow the prescription in~\cite{Cunningham1975}, where matter in the accretion disk moves in Keplerian orbits outside the innermost stable circular orbit (ISCO; $r=6M$) and infalls with conserved quantities equal to those at the ISCO when at radii inside the ISCO. While we only consider one velocity profile in this work, others may be valid\footnote{For example, it is possible that the majority of emission intensity in M87$^*$ images is coming from the jet~\citep{2023ApJ...950...38E,Chang24}. In that instance, the velocity profile could be very different, and this model would be ill-suited. However, under the assumption above that the majority of emission comes from a disk-like accretion profile, assuming matter has some type of orbital velocity profile is a reasonable approximation.}. For example, MAD GRMHD simulations typically observe sub-Keplerian orbits~\citep[see e.g.][]{Chael21,LiaAsym22,Conroy23}. As such, the results reported below are contingent on the choice of velocity profile and may change if other profiles are considered. For instance, \citet{Pu2018} and~\citet{Vincent22} both use analytic models varying black hole spin, disk thickness, and disk velocity to assess how these parameters change image and visibility domain features. Similarly, \cite{LiaAsym22} demonstrate how different disk parameters influence the velocity profile, and quantify the effect these have on the brightness asymmetry in black hole images. In Appendix~\ref{sec:SpinInvestigation}, we examine how changing the velocity profile affects some image features we use in the rest of this analysis.

\begin{figure}
  \includegraphics[width=\linewidth]{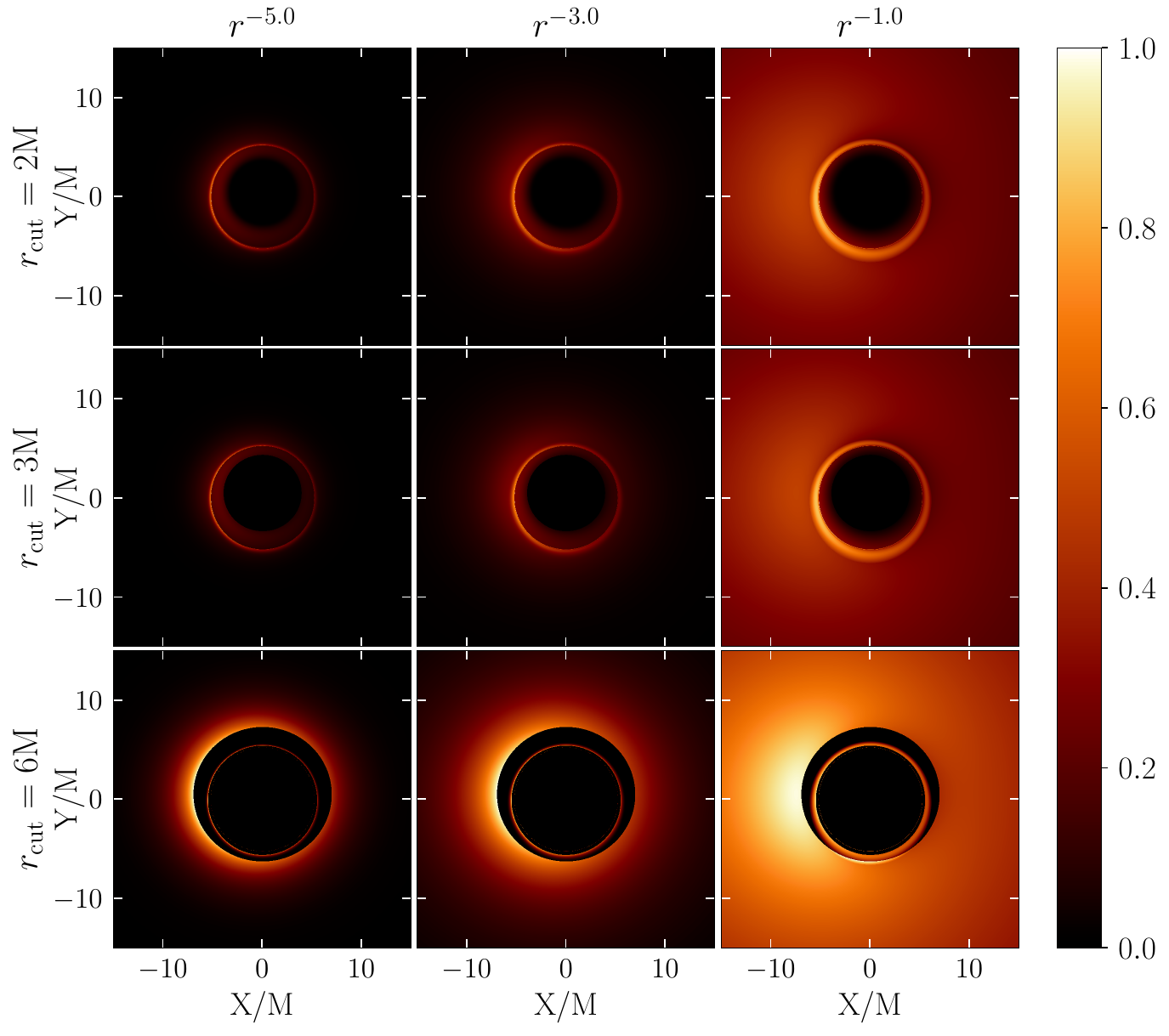}
\caption{Theoretical observational images of a Schwarzschild black hole with an accretion disk inclined at an angle $\theta_0 = 17^\circ$. The emission profile is given by Equation~\ref{eq:Ir} for cases $\beta = \{-5,-3,-1\}$ and $r_{\rm cut} = \{2,3,6\}$M. Notice the left-right brightness asymmetry due to the Doppler effect and the top-bottom geometric asymmetry due to the inclination. The lensing and photon rings are very narrow. When visualizing our images, we use EHT's perceptually uniform colormap \texttt{afmhot\_us} from the \texttt{ehtplot} library~\citep{ehtplotCode}.}
\label{fig:imageGrids}
\end{figure}

Since we assume a geometrically thin disk, we can estimate the total intensity that a distant observer sees as the sum of intensities at each intersection point given by the transfer functions:
\begin{eqnarray}
I_{\nu^\prime}(b,\alpha) = \sum_{i = 1}^{3}\left[g(b,\alpha,r_i)\right]^3 I_{\nu}(r_i(b,\alpha)) \; ,
\label{eq:sum}
\end{eqnarray}
where $r_i$ is the radius corresponding to the $i$-th intersection of a ray traced back from the image plane coordinate $(b,\alpha)$ to the disk. The EHT images of M87* are reported in units of brightness temperature, which is defined to be directly proportional to the specific intensity~\citepalias{EHT4}. Therefore, we can directly compare theoretical images created from Equation~\ref{eq:sum} to EHT's images, modulo an overall brightness re-scaling.

Figure~\ref{fig:imageGrids} plots theoretical images for three fiducial emission profile cutoffs, $r_{\rm cut} = {\{2,3,6\}}M$, and power laws, $\beta =\{-1,-3,-5\}$. We find that in the cases of $r_{\rm cut} = 3M$ and $r_{\rm cut} = 6M$, there exists a sharp cutoff at $b \sim r_{\rm cut} + 1M$, as expected by the ``just add one'' approximation~\citep{1910,2107}. In the $r_{\rm cut} = 2M$ case, the direct emission decays slowly to the horizon. We also see the ``lensing ring'' and ``photon ring'' as defined in \S\ref{sec:theoryBend}, which exist within the brightness depression in the $r_{\rm cut} = 6M$ case. In the $r_{\rm cut} = 3M$ and $r_{\rm cut} = 2M$ cases, the rings exist atop the direct emission. Finally, we see a left-right brightness asymmetry due to the Doppler shift, as well as a top-down geometric shift in the rings due to the inclination of the accretion disk.

\begin{figure}
\includegraphics[width=\linewidth]{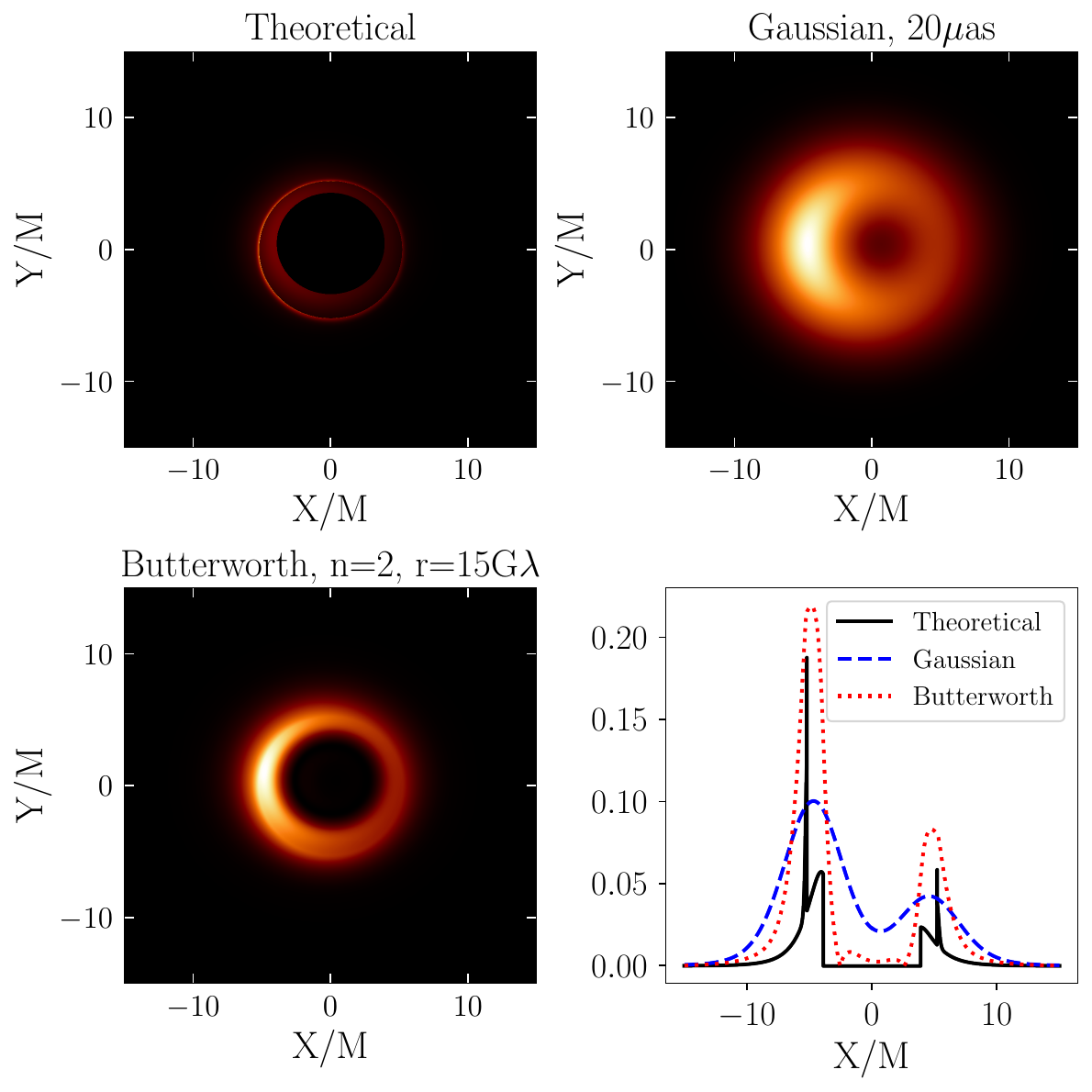}
\caption{(Top left) Theoretical image generated with a disk cutoff of $r_{\rm cut} = 3M$ and emission profile power law of $\beta = -6$. (Top right) Same image blurred with a Gaussian filter to the equivalent 20 $\mu \rm as$ resolution of the \texttt{DIFMAP} algorithm. (Bottom left) Same image instead blurred with a Butterworth filter to compare with the \texttt{PRIMO} algorithm. (Bottom right) Emission profile of a horizontal cross-section passing through the origin for the theoretical (black solid line), Gaussian-blurred (blue dashed line), and Butterworth-blurred (red dotted line) images. We see that the lensing ring and photon ring present in the theoretical image are now completely smoothed out with either filtering procedure.}
\label{fig:blur-types}
 \end{figure}

As our generated images in effect have infinite resolution, we apply a filter to our images to match the effective resolution of EHT array, in effect blurring our images. In this work, we consider two types of filters. The first is a Gaussian filter used to blur our theoretical images to the equivalent of the $20 \mu\rm as$ resolution of the image reconstruction in ~\citetalias{EHT4} using the \texttt{DIFMAP} imaging algorithm~\citepalias{1997ASPC..125...77S,2011ascl.soft03001S,EHT4}. The second is a Butterworth filter~\citep{Butterworth1930}, where we set $n = 2$ and $r = 15 \rm G \lambda$ to match with the \texttt{PRIMO} algorithm~\citep{2020arXiv200406210P,PRIMO23,PRIMOintro}\footnote{Note that, since the images we generate are dimensionless, we must take care to keep a constant effective resolution when using the filters to blur our theoretical images. In practice, we fix our resolution to be a percentage of our field of view such that increasing the mass of the black hole will increase the resolution of our images since the angular scale of the image parameters will increase with increasing black hole mass.}. Sample images with parameters $r_{\rm cut} = 3 \rm M$ and $\beta = -6$ with either the Gaussian filter or Butterworth filter applied are shown in the top right and bottom left panels of Figure~\ref{fig:blur-types}, respectively. A cross-section for the theoretical and blurred images are shown in the bottom right plot of Figure~\ref{fig:blur-types}. The cross-sections demonstrate that most of the sharp features, including the photon ring, are smoothed out and imperceptible in the blurred images. We also see that the Butterworth filter in general leads to thinner rings than the Gaussian filter. 

\section{\label{sec:imageFeatures}Image Analysis}
With the black hole images generated using our toy model from \S\ref{sec:theory}, we can now investigate robust image features to extract astrophysical properties of the black hole and accretion disk emission profile. Possible image features to examine include the diameter of the maximum brightness curve, $d$, the full width half maximum of the bright ring, $w$, the brightness asymmetry, $A$, and the ratio of minimum to maximum brightness, $f_c$. We use the definitions of these image features as specified in~\citetalias{EHT4}.

There are many astrophysical properties that will affect these features, such as the radial and azimuthal velocity profiles of the matter in the accretion disk, the emission profile $I(r_{\rm cut},\beta)$, the inclination of the disk, the mass of the black hole, and the spin of the black hole. In this work, we fix the black hole spin ($a = 0$, Schwarzschild), accretion disk inclination ($\theta_0 = 17^\circ$, estimated inclination of the M87* jet~\citep{2018ApJ...855..128W}), and disk velocity profile, and focus on how the mass and emission profile parameters change the image features, yielding three free parameters $(r_{\rm cut},\beta,M)$ to constrain.

\subsection{\label{sec:LengthScale} Calibrating theoretical images}
Using the formalism outlined in \S\ref{sec:theory}, we generate images for a range of emission profiles. Since black hole images scale directly with black hole mass, we simulate our theoretical images using $M=1M_\odot$ such that our theoretical images are effectively dimensionless. Then, we translate our dimensionless images to observed images using the angular scale corresponding to one gravitational radius, $\theta_g$, where $\theta_g = \frac{G M}{c^2 D}$, with black hole mass $M$ and distance to the black hole, $D$, with $D = 16.8$ Mpc to be consistent with \citetalias{EHT6}. As mentioned previously, when scaling our theoretical images with mass, we take care to keep a constant effective resolution in our blurred images. In practice, if we have some dimensionless image feature, such as diameter, we can relate this dimensionless diameter measurement to an angular-scale diameter measurement using
\begin{equation}
    f = \gamma_f \theta_g,
    \label{eq:GeneralCalib}
\end{equation}
where $f$ is the angular length of a given feature (such as diameter) and $\gamma_f$ is the dimensionless `calibration' parameter associated with that given feature. For a fixed distance to the black hole, changing the mass of the black hole changes the angular scale of the generated image features. In reality, if we measure the dimensionless image features in our generated images, $\gamma_f$, and are given the reported angular-scale image features from EHT's images, $f$, we can infer a mass for the black hole by rearranging Equation~\ref{eq:GeneralCalib} to find 
\begin{equation}
    M = \frac{c^2 D}{G} \frac{f}{\gamma_f} \; .
    \label{eq:massCal}
\end{equation}
Note that Equation~\ref{eq:GeneralCalib} is only used for image features that have an angular scale, such as the ring diameter $d$ and ring width $w$. Other image features that do not have an inherent angular scale, such as the brightness asymmetry, $A$, and brightness depression ratio, $f_c$, can be compared directly to physical values reported by EHT (i.e. $A_{\rm EHT} = \gamma_A$).

As a caveat, our theoretical images provide image features with full angular dependence such that
\begin{equation}
    f(\alpha) = \gamma_f(\alpha) \theta_g,
\end{equation}
where the origin can be defined as the location of the minimum brightness within the central brightness depression. Keeping the angular dependence in theoretical images would yield additional information about, for example, the inclination. While we average over image-plane angle $\alpha$ in this work to be consistent with current analyses~\citepalias{EHT4,PRIMO23}, investigation into the angular dependence of astrophysical properties of black holes may break degeneracies present in the angle-averaged image. For example, including the angular brightness asymmetry will constrain the black hole spin or accretion disk inclination~\citep[see e.g.][]{LiaAsym22}.

\subsection{\label{sec:ImageComparison} Comparing theoretical images to observed images}
We wish to compare multiple image features from our generated images at once to those of the observed images. We take the ratio of our two length-scale image features to measure the fractional width, $w/d$. Using $w/d$ scales out the mass-dependence and allows for a direct comparison between the theoretical and observed images. As mentioned previously, we can compare the brightness asymmetry, $A$, and central brightness depression ratio, $f_c$, directly. However, the observed images have uncertainties associated with the inferred image features. Table~\ref{tab:truths} reports the measurements with uncertainties that we use in this work when comparing to the \texttt{DIFMAP} and \texttt{PRIMO} images. \texttt{DIFMAP} values are calculated from the median measurements from all four days of observation during EHT's 2017 observing run~\citepalias{EHT4}. \texttt{PRIMO} values are taken directly from~\citet{PRIMO23}. To take uncertainties in the true features into consideration, we allow for up to one standard deviation in the measured values such that an image is considered viable if it satisfies
\begin{equation}
(\hat{f} - \Delta \hat{f}) \leq \gamma_{\hat{f}} \leq (\hat{f} + \Delta \hat{f}) \; ,
    \label{eq:1sigma}
\end{equation}
where $\hat{f}$ and $\gamma_{\hat{f}}$ now must be a ratio, and not a length-scale feature in the observed image and theoretical image, respectively. In the instance where there is only a confident upper bound reported for a feature, we consider our theoretical model viable if it satisfies
\begin{equation}
\gamma_{\hat{f}} \leq \hat{f}_{\rm max} \; ,
    \label{eq:upperbound}
\end{equation}
where $\hat{f}_{\rm max}$ is the confident upper bound in the observed image feature. If Equations~\ref{eq:1sigma} and~\ref{eq:upperbound} are satisfied for $w/d$, $A$, and $f_c$ simultaneously, we consider the theoretical model a viable combination of cutoff, $r_{\rm cut}$, and power law, $\beta$. If not, that emission profile is ruled out. Finally, for viable emission profiles, we use the measured diameter, $d \pm \Delta d$, and the `calibration' diameter, $\gamma_d$, from our theoretical images to find a posterior on mass for that emission profile.

\begin{table}[]
    \centering
    \begin{tabular}{|c|c|c|}
    \hline
       Feature  & \texttt{DIFMAP} & \texttt{PRIMO}  \\
       \Xhline{2\arrayrulewidth}
        $d$ [$\mu$as] & $40 \pm 3$ & $41.5 \pm 0.6$ \\
        \hline
        $w/d$ & $\leq 31/37$ & $ \leq 0.23$ \\
        \hline
        $A$ & $0.22 \pm 0.04$ & \mbox{--} \\
        \hline
        $f_c$ & $ \leq 0.45$ & \mbox{--} \\
        \hline
    \end{tabular}
    \caption{Measurements of the image diameter, $d$, fractional width, $w/d$, brightness asymmetry, $A$, and central brightness depression ratio, $f_c$, from \texttt{DIFMAP} and \texttt{PRIMO} used for our comparisons in this work. \texttt{DIFMAP} values are taken to be the median values calculated from all four observing days reported in Table 7 of~\citetalias{EHT4}. \texttt{PRIMO} measurements are taken from~\citet{PRIMO23}, and do not include measurements of the brightness asymmetry or central brightness depression ratio.}
    \label{tab:truths}
\end{table}

Note that the measurement errors we use correspond to the combined statistical and observational uncertainties across the four days of observation in the original 2017 data~\citepalias{EHT4}. These uncertainties, however, do not consider additional variations in the image feature measurements due to stochastic processes, such as accretion turbulence. As we neglect potential effects from inferring astrophysical properties from single snapshots in this work, we stress that the results we present below in \S\ref{sec:results} are thus optimistic estimates, and could be potentially influenced by turbulence (see e.g.~\citetalias{EHT6};~\citealp{2022ApJ...928...55P}).

\section{\label{sec:results} M87* Image comparison}
Using our toy model from \S\ref{sec:theory}, we generate a suite of theoretical black hole images varying the cutoff and power law of the accretion disk emission profile as well as the mass of the black hole. 
In the following subsections, we use the method outlined in \S\ref{sec:ImageComparison} to compare our generated filtered images to the observed images given by the original EHT data~\citepalias{EHT4} and the reanalyzed images from~\cite{PRIMO23}.

\subsection{\label{sec:M87-Gauss} Comparison to the original EHT image}

We note that EHT used three separate methods to analyze their raw 2017 data~\citepalias{EHT4, EHT6}. These methods include image reconstruction with imaging algorithms, fitting geometric models in the visibility domain, and fitting GRMHD snapshots to visibility data. Since the analysis we develop in this paper only considers the image-domain representation of the features, it is only consistent to compare our theoretical images to the data provided using the imaging algorithms described in~\citetalias{EHT4}, and as such we will only make use of the reported image features in the mentioned paper. Furthermore, each imaging algorithm presented in~\citetalias{EHT4} uses different smoothing conditions. We compare our images to those created using the \texttt{DIFMAP} algorithm~\citep{1997ASPC..125...77S,2011ascl.soft03001S} with an effective resolution of $20\mu \rm as$, although one could extend these results comparing to the \texttt{eht-imaging}~\citep{ehtimaging1,ehtimaging2,ehtimagingZenodo} and \texttt{SMILI}~\citep{SMILI1,SMILI2,SMILIzenodo} algorithms presented in~\citetalias{EHT4} as well. Although \texttt{eht-imaging} and \texttt{SMILI} have smaller reported widths of the bright ring due to their higher resolution, we expect that these would only lead to slightly different upper bounds on the emission profile power law $\beta$ (see discussion on power law constraints in \S\ref{sec:PRIMOresults}).

\begin{figure}
    \centering
    \includegraphics[width=\linewidth]{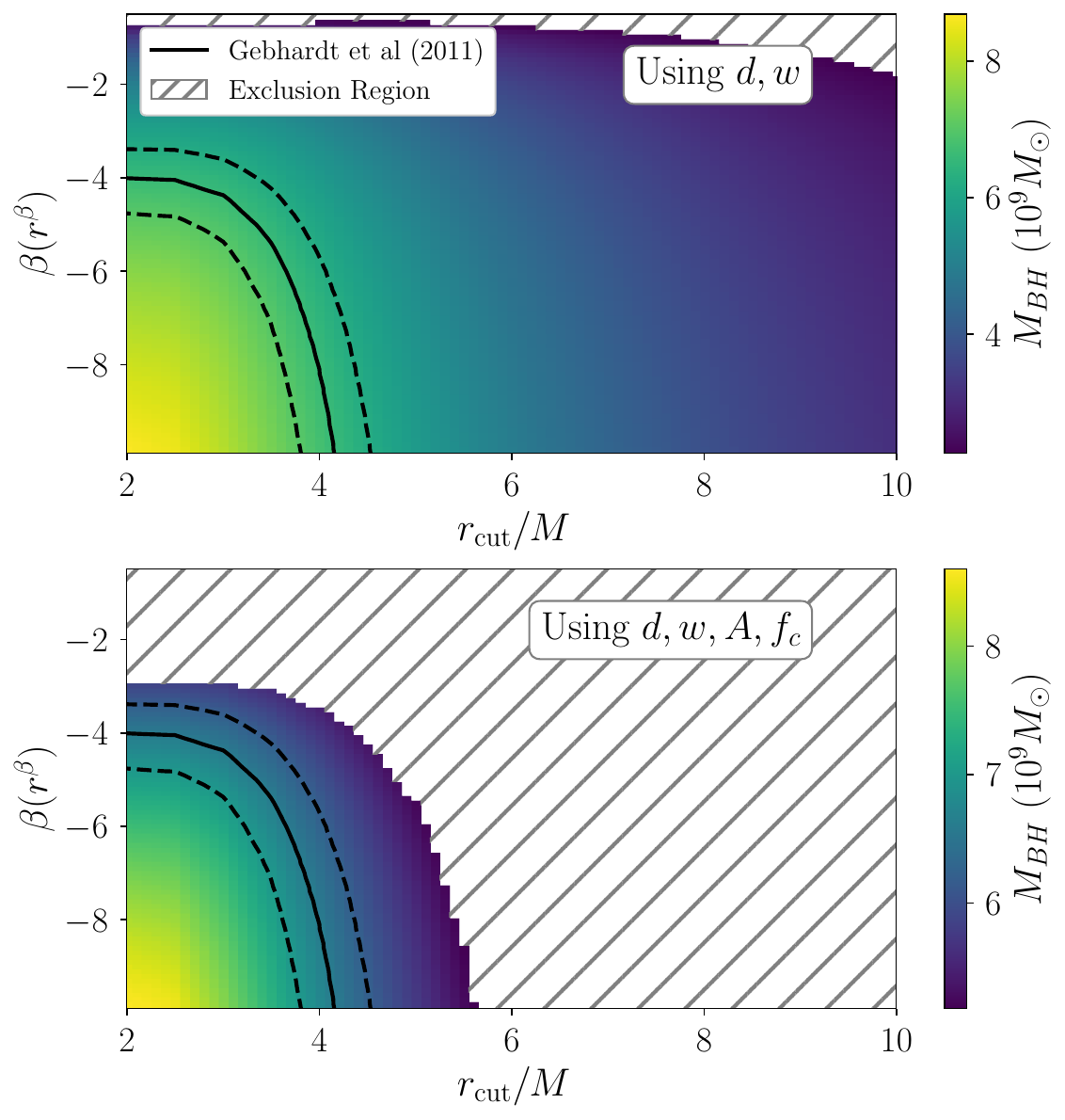}
    \caption{(Top) Median inferred masses for each emission profile to be consistent with EHT's image for M87$^*$ when comparing against the reported diameter and width observations of the original EHT analysis. Emission profile combinations that are not viable are indicated by the exclusion region (gray hatched region). (Bottom) Same as the top panel, but now additionally considering constraints on the brightness asymmetry and central brightness depression ratio. Marginalizing over $\beta$ and $r_{\rm cut}$ yields a mass estimate of $6.6^{+1.2}_{-1.0}\times 10^9 M_\odot$ to 68$\%$ confidence in the constraint region when considering all four image features. We include contours from the stellar-dynamics mass estimate of $(6.6\pm0.4)\times 10^9 M_\odot$ for M87* from~\cite{StellarDyn}, demonstrating that the stellar-dynamics mass estimate is consistent with the viable emission profiles from analyzing the EHT image features.}
    \label{fig:DIFMAPcontours}
\end{figure}

Figure~\ref{fig:DIFMAPcontours} depicts the median masses inferred using Equation~\ref{eq:massCal} for each emission profile that is consistent with EHT's reported diameter and width (top panel), and including brightness asymmetry and central brightness depression ratio (bottom panel) simultaneously. Note that more emission profiles may be viable outside of the $(r_{\rm cut},\beta)$ region we chose to investigate here. Since we are agnostic to the emission profile, all masses for M87$^*$ in the range of recovered viable emission profiles are allowed. However, if we marginalize over $r_{\rm cut}$ and $\beta$, we find the mass of M87* to be $4.3^{+2.2}_{-1.2}\times 10^9 M_\odot$ to 68$\%$ confidence within the constraint region using only the diameter and fractional width, and restricting the prior range to $r_{\rm cut} \in [2,10]M$ and $\beta \in [-10,-1/2]$. We also find that the emission profile must have a power law of $\beta \leq -0.7$. When instead considering all four image features, we find a new mass range of $6.6^{+1.2}_{-1.0}\times 10^9 M_\odot$ within the constraint region, and emission profiles must have a $\beta \leq -3$ and $r_{\rm cut} \leq 5.6M$. In Figure~\ref{fig:DIFMAPcontours}, we also include a contour (black solid and dashed lines) for the 68$\%$ credible mass range reported for M87* from stellar dynamics~\citep{StellarDyn} as a reference. We find the stellar dynamics mass estimate to be consistent with our analysis using EHT image features and our simplified accretion disk model.

Although considering brightness features (i.e., brightness asymmetry and central brightness depression ratio) further constrains the range of viable emission profiles and masses, they do not completely break the degeneracy between the mass, cutoff, and power law due to the large uncertainties in the observed features. Nonetheless, some constraints on viable emission profiles are evident. First, including an upper bound on the width of the bright ring almost directly corresponds to an upper bound on the power law, $\beta$. Likewise, a broad power law will wash out the brightness asymmetry, and thus measuring an asymmetry further rules out broad $\beta$. Similarly, measuring a brightness asymmetry leads to an upper bound on the accretion disk cutoff, as a large cutoff reduces the Doppler redshift, leading to a less pronounced brightness asymmetry. One of our main findings is that EHT images do not require emission from the accretion disk to extend all the way to the horizon.

Astrophysical processes such as the accretion disk velocity profile would be expected to change image features, including the brightness asymmetry\footnote{Recall that we assume the intrisic emission is symmetric, so any intrinsic brightness asymmetry in the emission would also affect these findings as well.}. While we do not consider any robust estimates of these variations, we explore how image features change with spin and disk velocity profile for a fiducial set of images in Appendix~\ref{sec:SpinInvestigation}. We find that the brightness asymmetry is sensitive to the choice of disk velocity profile and black hole spin. As such all constraints found using brightness asymmetry above are conditional on our assumption of Keplerian orbits outside the ISCO and may change as further astrophysics is incorporated.

\subsection{\label{sec:PRIMOresults} Comparison to \texttt{PRIMO} results}

\begin{figure}[tbh]
    \centering
    \includegraphics[width=\linewidth]{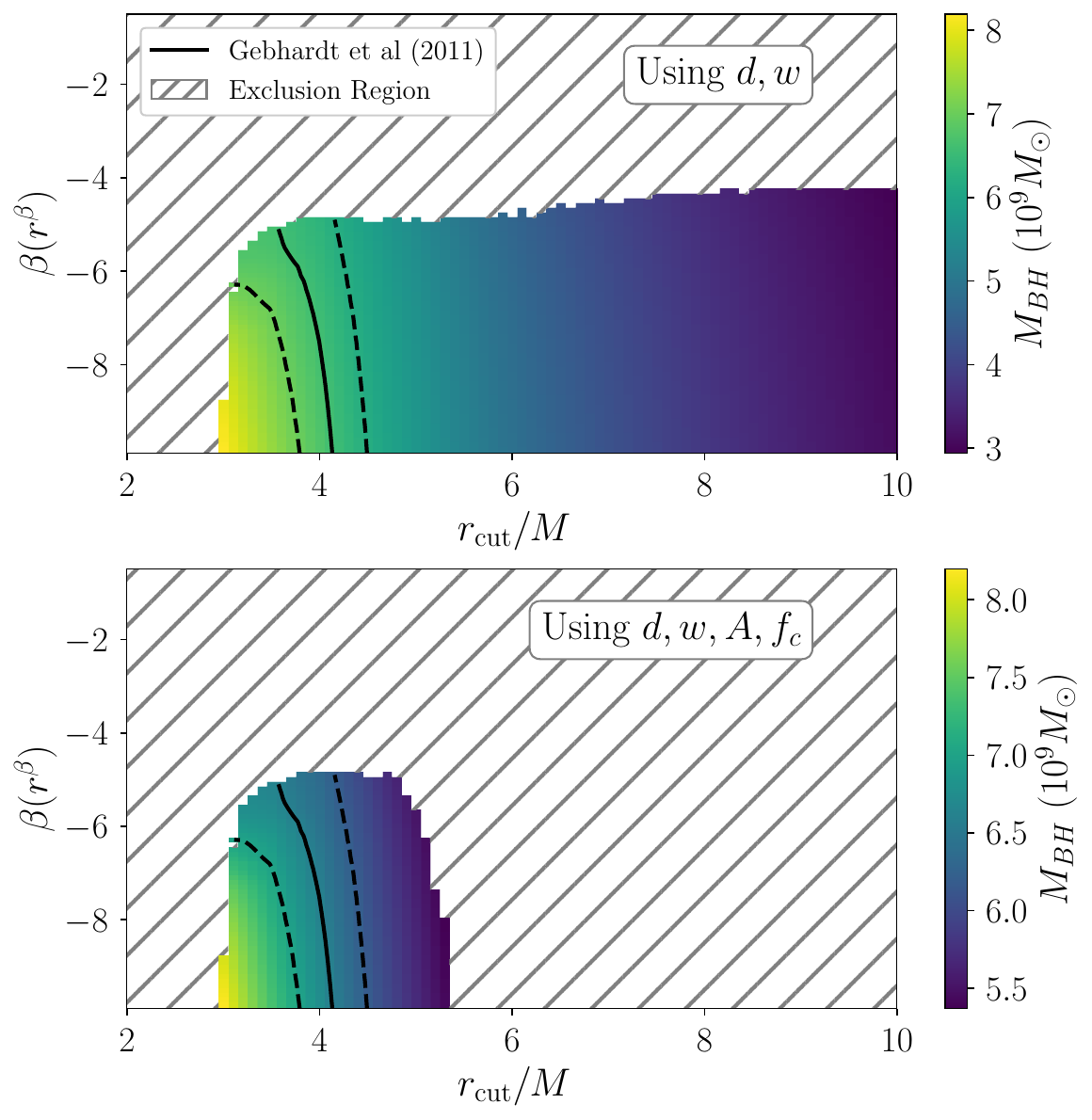}
    \caption{(Top) Median inferred masses for each emission profile to be consistent with the reported diameter and width from the \texttt{PRIMO} image analysis for M87$^*$. Emission profile combinations that are not viable are indicated by the exclusion region (gray hatched region). (Bottom) Same as the top panel, but additionally incorporating the reported brightness asymmetry and central brightness depression ratio measurements from the \texttt{DIFMAP} algorithm used in the original EHT analysis. We also include contours from the stellar-dynamics estimate of $(6.6\pm0.4)\times 10^9 M_\odot$ for M87* from~\cite{StellarDyn} (black solid and dashed lines). When marginalizing over $r_{\rm cut}$ and $\beta$ and considering all four image features, we find a mass estimate of $6.4^{+0.7}_{-0.7}\times 10^9 M_\odot$ to 68$\%$ confidence inside the constraint region. Notice that \texttt{PRIMO}'s images imply that the emission is cut off at $\geq 3M$ regardless of brightness constraints, and does not continue all the way to the event horizon. 
    }
    \label{fig:PRIMOcontours}
\end{figure}

The results from \S\ref{sec:M87-Gauss} consider the original analysis from the EHT collaboration. Here, we instead use a Butterworth filter to compare out theoretical images to the image reconstructed with \texttt{PRIMO} as presented in~\citet{PRIMO23}. Using a Butterworth filter keeps more information at the baselines to which EHT is sensitive to~\citep{2020arXiv200406210P}. However, since \texttt{PRIMO} uses GRMHD simulations in its training set, the results we present below are less agnostic to GRMHD than the results found in \S\ref{sec:M87-Gauss}. 

Figure~\ref{fig:PRIMOcontours} demonstrates the viable emission profiles when using a Butterworth filter to blur theoretical images. Since \cite{PRIMO23} only reports a diameter and confident upper bound for the width of the bright ring, we first only compare these two image features to our theoretical images (see the top panel in Figure~\ref{fig:PRIMOcontours}). An immediate difference from the previous results is that, since \texttt{PRIMO}'s new smoothing procedure leads to thinner rings, the emission profile power law is constrained to be steeper, around $\beta \leq -4.3$. Another evident characteristic is that the event horizon is now ruled out with the currently reported \texttt{PRIMO} results, with viable emission profiles requiring $r_{\rm cut} \geq 3M$. This exclusion is due to the smaller upper bound for the fractional width, $w/d \lesssim 0.23$. As the cutoff moves from $3M$ to $2M$, the diameter will remain approximately the same due to gravitational redshifting of the emission close to the horizon. On the other hand, the width of the ring increases due to the additional emission extending within $3M$, thereby increasing the ratio of ring width to diameter. Due to these new bounds on the emission profile cutoff and power law, the mass range is now constrained to be $4.3^{+2.0}_{-1.0}\times 10^9 M_\odot$ to 68$\%$ confidence within the constraint region, when restricting the prior range to $r_{\rm cut} \in [2,10]M$ and $\beta \in [-10,-1/2]$.

Although \citet{PRIMO23} do not report a brightness asymmetry or central brightness depression ratio, as an example of how these features would influence the above results, we use the reported values for brightness asymmetry and central brightness depression ratio from the  \texttt{DIFMAP} image reconstructions in~\citetalias{EHT4}. Including these two features in the analysis provides contours shown in the bottom panel of Figure~\ref{fig:PRIMOcontours}. Although there are only marginally tighter constraints on the power law ($\beta \leq -4.9$), including information on the brightness asymmetry further restricts the viable emission profile cutoffs to be between $(3 \mbox{--} 5.3) M$ and the mass of M87* to be $6.4^{+0.7}_{-0.7}\times 10^9 M_\odot$ when marginalizing over $r_{\rm cut}$ and $\beta$. We again include a contour (black solid and dashed lines) for the 68$\%$ credible interval in the stellar-dynamics mass estimate from~\cite{StellarDyn} to demonstrate that our results are consistent with independent measurements of mass.

\section{\label{sec:implications} Implications for M87*}
In \S\ref{sec:M87-Gauss} and \S\ref{sec:PRIMOresults} we compare theoretical black hole images to the features reported from two imaging algorithms, \texttt{DIFMAP} and \texttt{PRIMO}, using EHT data of M87*. Since the \texttt{DIFMAP} algorithm has a resolution of 20$\mu\rm as$, our above analysis is mainly limited by the uncertainty in the width of the bright ring. We find when comparing our images to the \texttt{DIFMAP} results that accretion disk emission may extend all the way to the event horizon ($r_{\rm cut}=2M$), although other cutoffs ($r_{\rm cut} \leq 5.6M$) outside the horizon are not ruled out. The mass of M87* is constrained to be within $ 6.6^{+1.2}_{-1.0}\times 10^9 M_\odot$ when using all four image features, consistent with the currently reported mass of $\sim 6.6\times 10^9 M_\odot$ from stellar dynamics~\citep{StellarDyn} and the $\sim 6.5\times 10^9 M_\odot$ reported mass from EHT~\citepalias{EHT6}. When instead comparing to the \texttt{PRIMO} results presented in~\cite{PRIMO23}, we now find that emission extending all the way to the horizon is ruled out, with the inner edge of the accretion disk required to be at or above $3M$. The mass of M87* is constrained to be within $6.4^{+0.7}_{-0.7}\times 10^9 M_\odot$ when additionally considering brightness asymmetry constraints from the \texttt{DIFMAP} reconstructions, still consistent with previous mass estimates.

If we instead wish to constrain the disk's inner cutoff, $r_{\rm cut}$, we can marginalize over the emission profile power law and the black hole mass. As we have shown in Appendix~\ref{sec:SpinInvestigation}, brightness asymmetry is conditional on the disk velocity profile and black hole spin used in this analysis. Therefore, we now use our constraints when only comparing to the diameter and width of the images, and instead additionally assume the mass of the black hole must be consistent with the $68\%$ credible interval from~\cite{StellarDyn}. Under these assumptions and using our above constraints comparing to the \texttt{PRIMO} results, we find that the disk inner edge will be around $r_{\rm cut} = 4.0^{+0.3}_{-0.3}M$ to 68$\%$ confidence within our viable emission profile constraint region.

\begin{figure}
    \centering
\includegraphics[width=\linewidth]{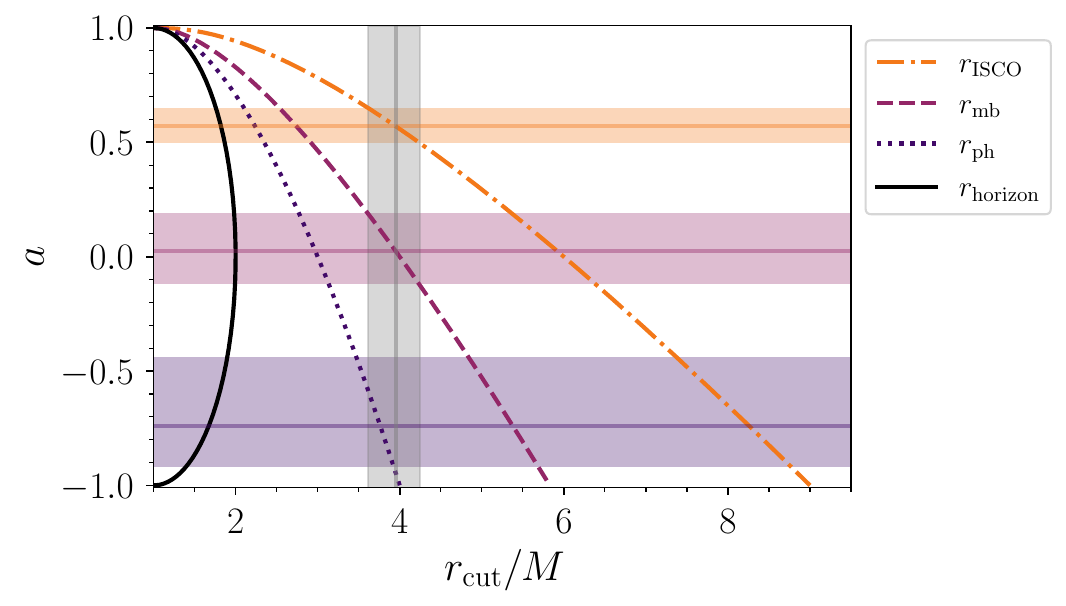}
    \caption{Estimate of the dimensionless black hole spin $a$ if we assume the cutoff $r_{\rm cut}$ corresponds to either the photon ring ($r_{\rm ph}$; purple shaded region), the marginally-bound orbit ($r_{\rm mb}$; pink shaded region), or the ISCO ($r_{\rm ISCO}$; orange shaded region) using the constraint on the disk inner cutoff inferred using the \texttt{PRIMO} algorithm while assuming the mass estimate from \cite{StellarDyn} (grey vertical shaded region). Note that in our conventions, $a>0$ indicates the accretion disk is prograde with respect to the black hole spin, and $a < 0$ indicates the disk is retrograde.}
    \label{fig:spinEst}
\end{figure}

In geometrically thin disks, it has been shown that the inner edge of the accretion flow coincides with the ISCO~\citep{NovikovThorne,2000astro.ph..4129P,2010MNRAS.408..752P}, and hence should be at $r = 6M$ for a Schwarzschild black hole. Figure~\ref{fig:spinEst} shows that this is inconsistent with the \texttt{PRIMO} image for Schwarzschild, and instead argues for a non-zero black hole spin with $r_{\rm cut} = 4.0^{+0.3}_{-0.3}M \equiv r_{\rm ISCO}$ (see further discussion about black hole spin below). Alternatively, in the case of \textit{slim} disks~\citep[see e.g.][]{1988ApJ...332..646A}, the inner edge no longer necessarily corresponds to the ISCO, as demonstrated in~\cite{2010A&A...521A..15A}. Similarly, Figures~1 and ~2 of~\citet{2011A&A...532A..41S} demonstrates that the flux and disk height can significantly decrease inside the ISCO for certain accretion rates. The cutoff we introduce in this work is most similar to the innermost radius with significant luminosity, $r_{\rm rad}$, introduced in~\cite{2010A&A...521A..15A}, which is shown to occur outside the horizon assuming a~\cite{NovikovThorne} flux radial profile. From Figure 13 in~\cite{2010A&A...521A..15A}, we see the inner edge depends on the accretion rate, but that an $r_{\rm rad} \equiv r_{\rm cut}  = 4.0^{+0.3}_{-0.3}M$ is reasonable for a mildly sub-Eddington accretion rate. Therefore, we may be able to discern information about the accretion rate around M87* by connecting $r_{\rm rad} \sim r_{\rm cut}$ to the accretion rate.

Alternatively, we can attempt to connect the cutoff $r_{\rm cut}$ to an intuitive physical radii such as the photon ring ($r_{\rm ph}$), the ISCO ($r_{\rm ISCO}$), or the innermost marginally-bound orbit ($r_{\rm mb}$)\footnote{For example,~\cite{2023ApJ...942...47Y} argue that the maximum emission in black hole images is very closely related to the black hole shadow, in which case we would take $r_{\rm cut}\sim r_{\rm ph}$ such that the central brightness depression in the generated image is located at the critical curve $b_c$~\citep[see e.g.][]{GHW}.}. If we take the stellar-dynamics mass estimate from~\cite{StellarDyn} and our constraints from \texttt{PRIMO}, we restrict the cutoff to $r_{\rm cut}  = 4.0^{+0.3}_{-0.3}M$. We can then translate this constraint in cutoff to a constraint on black hole spin. For example, if we assume the cutoff corresponds to the photon ring ($r_{\rm ph}$), then the black hole must be spinning with dimensionless spin $a = -0.7^{+0.3}_{-0.2}$ (where $a<0$ indicates a retrograde orbit) to align with this assumption to 68$\%$ confidence in the constraint region. Likewise, if we assume the disk \textit{must} cut off at the ISCO (i.e. $r_{\rm cut} \equiv r_{\rm ISCO})$, the black hole must have a spin of $a = 0.6^{+0.1}_{-0.1}$. Finally, if we assert $r_{\rm cut} \equiv r_{\rm mb}$, then the spin of the black hole must be $a = 0.02^{+0.17}_{-0.15}$. These estimates are shown for the three physical radii chosen, $r_{\rm ph}$, $r_{\rm mb}$, and $r_{\rm ISCO}$, as the purple, pink, and orange shaded regions in Figure~\ref{fig:spinEst}, respectively. However, we stress that these spin estimates rely on the assumption of the nature of the cutoff ($r_{\rm ph}, r_{\rm mb}, r_{\rm ISCO}$), and as such this is a toy-model demonstration of how one might estimate the spin of the black hole without explicitly incorporating the Kerr metric. Future analyses will explicitly incorporate Kerr black holes, and thus provide more robust constraints on spin.

While we have employed a highly simplified toy model for the accretion disk emission profile, we argue that a more sophisticated description of the accretion disk would generally be expected to find consistent conclusions. Regardless of the algorithm, we find that the mass of M87* should conservatively be within $(5.6\mbox{--}7.8)\times 10^9 M_\odot$ when considering all four image features considered in this work, the accretion disk brightness must drop off with a power law $\beta \leq -3$, and disk cutoffs outside the event horizon are viable, and in the case of \texttt{PRIMO} constraints, required. Therefore, one cannot conclude that the central brightness depression in the center of EHT images directly corresponds to the event horizon or black hole shadow. Instead, the lack of observed emission at the center may be due to a lack of actual emission from the accretion disk itself, rather than a shadowing effect.

\section{\label{sec:discConc} Discussion and conclusion}
We have introduced a simple two-parameter toy model for the emission profile of a geometrically and optically thin accretion disk orbiting a black hole. We use this toy model to create theoretical images which we compare to EHT observations using the \texttt{DIFMAP} or \texttt{PRIMO} imaging algorithms, respectively. We then relate the dimensionless and angular-scale image features to infer the mass of M87* and viable emission profile parameters. We find that current black hole images are insufficiently resolved to precisely constrain both the mass of the black hole and a two-parameter toy model for the emission profile of the accretion disk. To reproduce the image features reported using the \texttt{DIFMAP} algorithm for diameter, width, brightness asymmetry, and minimum brightness depression ratio, we require the mass of M87$^*$ to be $6.6^{+1.2}_{-1.0}\times 10^9 M_\odot$ to 68$\%$ confidence within the range of viable emission profiles found in \S\ref{sec:M87-Gauss}, and the cutoff and power law for our toy model emission profile to be less than $r_{\rm cut} \leq 5.6M$ and $\beta \leq -3$. If we instead compare to reported diameter and fractional width values from the \texttt{PRIMO} algorithm, we find M87* has a $68\%$ credible mass range between $4.3^{+2.0}_{-1.0}\times 10^9 M_\odot$ when restricting our parameter space to $r_{\rm cut} < 10M$ and $\beta > -10$. We further infer that cutoffs are restricted to be greater than $3M$ and the power law is restricted to be $\beta \leq -4.3$. If we use the brightness asymmetry and central brightness depression ratio reported by \texttt{DIFMAP}, combined with the diameter and width reported by \texttt{PRIMO}, we find the cutoff to be between $(3\mbox{--}5.3)M$ with a 68$\%$ credible mass range of $6.4^{+0.7}_{-0.7} \times 10^9 M_\odot$. This suggests that the accretion disk emission drops off before reaching the horizon, leading to an observed central brightness depression, but note that this does not necessarily correspond to a black hole shadow.

This analysis has focused on representations of EHT data in the image domain. However, it is more robust to analyze images in the visibility domain, as this would allow direct comparison with EHT's data and would minimize uncertainties due to the fitting algorithms. We leave this for future work.
 
We restrict our analysis to emission profiles for geometrically and optically thin accretion disks. Although adding in a geometrically thick disk may not significantly change the results according to the `equatorial approximation'~\citep{Gralla2020}, having a non-zero scale height has been shown to influence the amount of emission inside the central brightness depression~\citep{2022ApJ...941...88O,Vincent22}. Furthermore, as accretion disks are often expected to be geometrically thick~\citepalias{EHT5}, incorporating such disks would improve the realism of our toy model. This would introduce another fitting parameter, the disk thickness, to constrain.

As mentioned above, there are additional factors such as inclination, disk velocity profiles, and black hole spin that will impact features in the image or visibility spectrum. Expanding the parameter space to incorporate these factors will lead to more robust astrophysical constraints. Considering the full intensity angular profile may further aid in breaking degeneracies that may arise between different astrophysical parameters. However, even with current EHT resolution, our computationally-inexpensive toy model captures many of the key physical features present in black hole images, confidently excludes interesting regions of parameter space, provides robust constraints on the black hole mass and accretion disk emission profiles, and may inform observables for future GRMHD simulations.

\section*{Acknowledgements}
We are grateful to Lia Medeiros, Samuel Gralla,
and Charles Gammie for extremely helpful and insightful discussions. DEH and AGG were supported by NSF grants  {PHY-2110507 and PHY-2513312}. DEH was also supported by the NSF-Simons AI-Institute for the Sky (SkAI) via grants NSF AST-2421845 and Simons Foundation MPS-AI-00010513, and by the Simons Collaboration on Black Holes and Strong Gravity through grant SFI-MPS-BH-00012593-07. AGG is grateful for the generous support of the Brinson Foundation and the ARCS Foundation, Illinois Chapter.

\software{numpy~\citep{2020Natur.585..357H}, scipy~\citep{2020NatMe..17..261V}, matplotlib~\citep{2007CSE.....9...90H}, jupyter~\citep{2016ppap.book...87K}, pandas~\citep{mckinney-proc-scipy-2010}, ehtplot~\citep{ehtplotCode}, AART~\citep{AARTcode}.}

\appendix

\twocolumngrid 

\section{\label{sec:SpinInvestigation} Investigating the effect of spin and velocity profile on image features}
\begin{figure}[h!]
    \centering
\includegraphics[width=\linewidth]{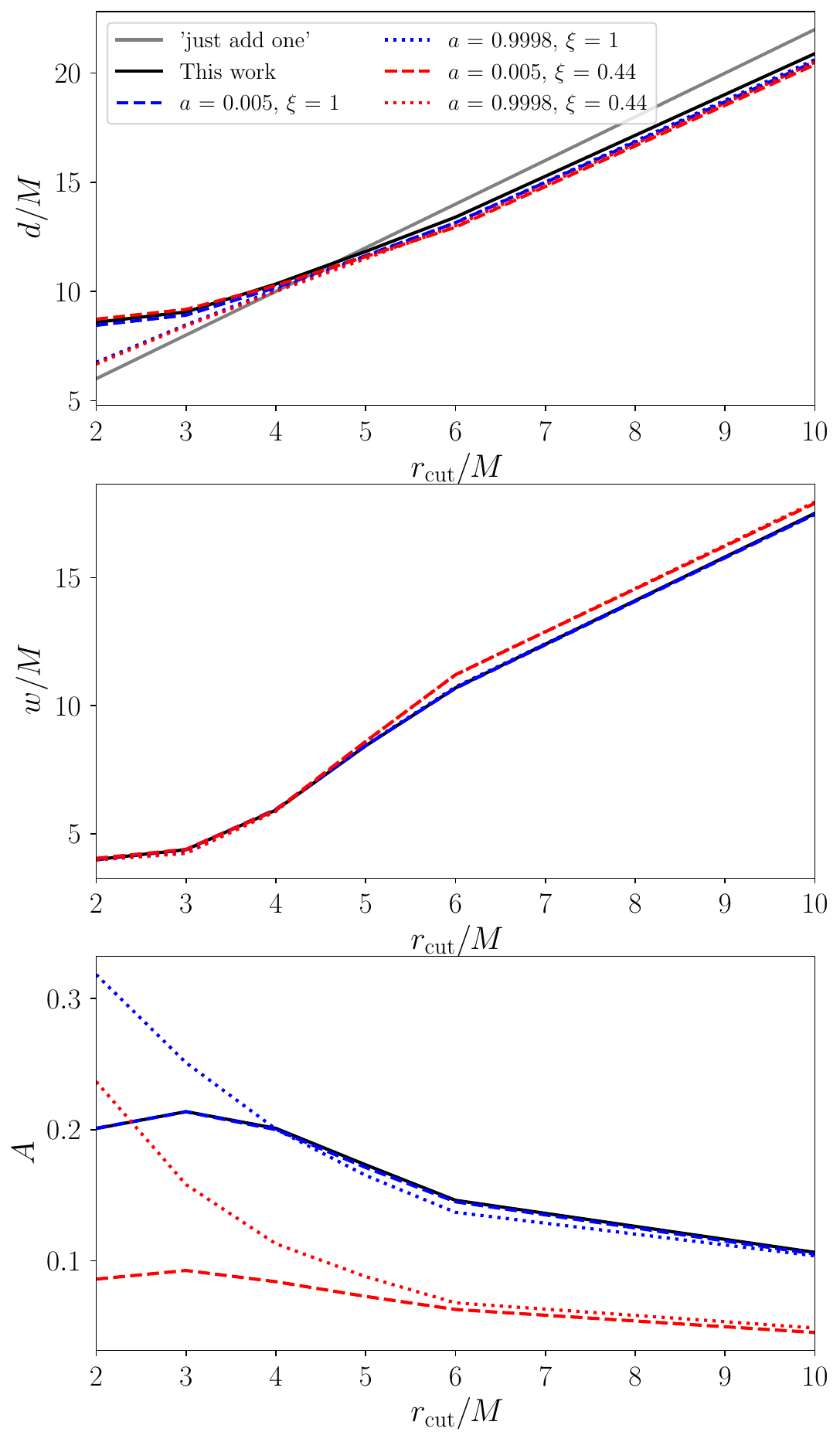}
    \caption{Measured diameter (top), width (middle), and brightness asymmetry (bottom) from images generated with the \texttt{AART} image generation code when varying the accretion disk inner cutoff, $r_{\rm cut}$, for two fiducial black hole spins, $a = 0.005$ (dashed lines) and $a=0.9998$ (dotted lines), and sub-Keplerianity parameters, $\xi=1$ (blue lines) and $\xi=0.44$ (red lines). We use a fiducial emission profile power law $\beta=-8$, with a black hole mass that yields a viable emission profile when compared to EHT values (see Figure~\ref{fig:DIFMAPcontours}). We also compare the image features measured for images generated in our analysis (black solid lines). The gray solid line in the top panel indicates what diameter would be inferred simply using the `just add one' approximation. 
    }
    \label{fig:A-vs-chi}
\end{figure}

In our analysis we have only considered Schwarzschild black holes with accretion disks that move on Keplerian orbits outside the ISCO and infall with conserved quantities within the ISCO. However, we expect image features to depend on the spin of the black hole and disk velocity profile. To investigate the strength of these variations, we compare the image features measured in this analysis to those measured when generating black hole images using the \texttt{AART} code~\citep{AARTpaper, AARTcode}, varying spin ($a = 0.005, 0.9998$) and the sub-Keplerianity parameter ($\xi = 0.44,1$), a factor that controls the disk angular momentum~\citep{AARTpaper}. We blur the images generated with \texttt{AART} with a Gaussian filter equivalent to the 20$\mu$as resolution from EHT's image reconstructions using the \texttt{DIFMAP} imaging algorithm. We set the power law of the emission profile to $\beta = -8$, which is always viable for both our blurring prescriptions (see Figs. \ref{fig:DIFMAPcontours} and \ref{fig:PRIMOcontours}). We show the image features measured as a function of cutoff in Figure~\ref{fig:A-vs-chi}, where we choose the mass to be such that the power law and cutoff are viable emission profiles as reported in Figure~\ref{fig:DIFMAPcontours}. Note that our features (solid black line) should closely match the blue dashed line, corresponding to the \texttt{AART} images with $a=0.005$ and $\xi = 1$. We find in Figure~\ref{fig:A-vs-chi} that this is true up to $\lesssim 1 \%$ for all cutoffs considered. 

A few trends are apparent. First, the width of the bright ring does not depend noticeably on black hole spin or accretion disk velocity profile for the range of cutoffs we consider, most likely due to the 20 $\mu$as Gaussian filter dominating the total width. However, while the velocity profile does not affect the diameter of the bright ring, increasing the black hole spin leads to smaller diameters at small disk cutoffs. Since the horizon for a Schwarzschild black hole is at $r=2M$, images are redshifted near this horizon, leading to an observed diameter that asymptotes near the horizon. However, near-extremal Kerr black holes have a horizon near $r = 1M$, and thus a disk cutting off near $r_{\rm cut}=2M$ will not be as substantially redshifted, leading to a smaller observed diameter of the image ring than the same disk orbiting a Schwarzschild black hole.

Finally, the brightness asymmetry of the bright ring depends both on the spin of the black hole and the velocity profile of the accretion disk. The spin of the black hole only appreciably affects the brightness asymmetry starting at cutoffs inside $r_{\rm cut} \lesssim 4M$, as the cutoff nears the horizon. Increasing the black hole spin increases the brightness asymmetry in this regime due to frame-dragging. Similarly, having a sub-Keplerian velocity profile outside the ISCO (red dashed and dotted lines) results in a substantially smaller brightness asymmetry than a disk that has a Keplerian velocity profile (blue lines), due to smaller disk velocities decreasing the Doppler effect. 
Therefore, while the width is robust to changes both in black hole spin and disk velocity profile, the diameter is affected by spins for $r_{\rm cut} \lesssim 4M$, although the effect is only noticeable for larger spins. However, the brightness asymmetry does depend on black hole spin and even more so on the disk velocity profile, and thus conclusions that depend upon brightness asymmetry observations are especially sensitive to assumption about the disk velocity profile and black hole spin.


\bibliography{references}{}
\bibliographystyle{aasjournalv7}

\end{document}